# Charge transfer as the key parameter affecting color purity of TADF emitters

Ramin Ansari[1], Wenhao Shao[2], Seong-Jun Yoon[3], Jinsang Kim *,[1,2,3,4], John Kieffer *,[3]

[1]Department of Chemical Engineering, University of Michigan, Ann Arbor, USA

[2]Department of Chemistry, University of Michigan, Ann Arbor, USA

[3]Department of Materials Science and Engineering, University of Michigan, Ann Arbor, USA

[4]Macromolecular Science and Engineering, University of Michigan, Ann Arbor, USA

**ABSTRACT:** The key factors determining the emission bandwidth of thermally activated delayed fluorescence (TADF) are investigated by combining computational and experimental approaches. To achieve high internal quantum efficiencies (IQEs) in metal-free organic light emitting diode via TADF, the first triplet ($T_1$) to first singlet ($S_1$) reverse intersystem crossing (rISC) is promoted by configuring molecules in an electron donor-acceptor (D-A) alternation with a large dihedral angle, which results in a small energy gap ($\Delta E_{ST}$) between $S_1$ and $T_1$ levels. This allows for effective non-radiative up-conversion of triplet excitons to singlet excitons that fluoresce. However, this traditional molecular design of TADF results in broad emission spectral bands (full-width at half-maximum = 70-100 nm). Despite reports suggesting that suppressing the D-A dihedral rotation narrows the emission band, the origin of emission broadening remains elusive. Indeed, our results suggest that the intrinsic TADF emission bandwidth is primarily determined by the charge transfer character of the molecule, rather than its propensity for rotational motion, which offers a renewed perspective on the rational molecular design of organic emitters exhibiting sharp emission spectra.

## INTRODUCTION

In an organic light emitting diode (OLED), electrically injected charge carriers form singlet and triplet excitons at a 1:3 ratio.[1–4] To overcome the energy conversion limitation imposed by forbidden emissive transitions from triplet states,[5] the focus has been primarily on organometallic and metal-free organic phosphors, owing to their efficient photon generation from triplet excitons by means of the heavy atom effect.[6–9] As an alternative, thermally activated delayed fluorescence (TADF) has been extensively studied. In this approach, high internal quantum efficiencies (IQEs) without using metals are achieved by promoting the $T_1$ to $S_1$ reverse intersystem crossing (ISC) in organic molecules designed to exhibit small energy gaps between $S_1$ and $T_1$ levels ($\Delta E_{ST}$).[10–16] Small $\Delta E_{ST}$ (< 0.1 eV) allow for non-radiative triplet excitons to be up-converted to singlet excitons before decaying radiatively. Therefore, 100% IQE can theoretically be realized. The prevailing molecular design strategy to minimize $\Delta E_{ST}$ is to configure conjugated organic groups into electron donor-acceptor (D-A) alternation with a large dihedral angle (Figure 1). This causes, the highest occupied molecular orbital (HOMO) and the lowest unoccupied molecular orbital (LUMO) to be localized on the donor and acceptor moieties, respectively, leading to small HOMO and LUMO orbital overlap. $\Delta E_{ST}$ amounts to twice the electron exchange energy, $J$, which is determined by the extent of spatial overlap of HOMO and LUMO.[17]

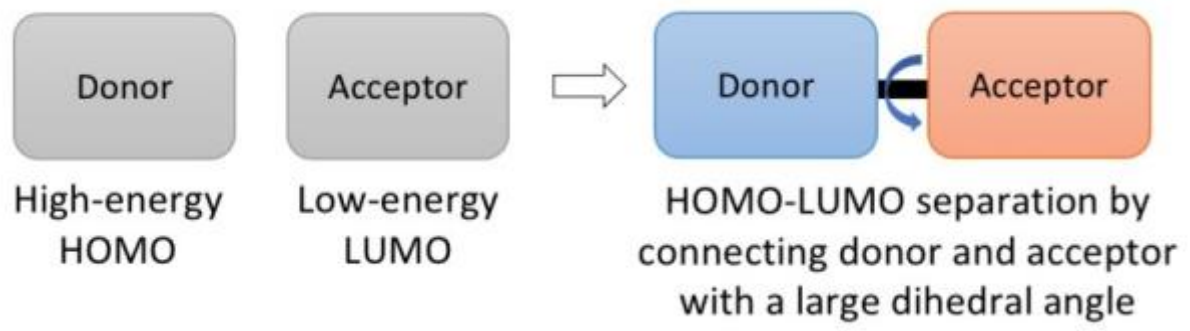

Figure 1. Conventional TADF molecular design. Donor and Acceptor groups can rotate around D-A connection which enhances the structural relaxation in the excited states, resulting in large stokes shift and a broad emission. However, in this contribution we systematically demonstrate that the molecular rotation is not the major cause of the broad emission of TADF.

Hence, the smaller the HOMO-LUMO overlap, the easier it is to achieve comparable energy levels for $S_1$ and $T_1$, resulting in small $\Delta E_{ST}$ values.[18,19] Accordingly, spatially separating HOMO and LUMO orbitals, leading to so-called charge transfer (CT) excited states, is perceived as essential for TADF, and constitutes the conventional design principle for such emitters. Unfortunately, it is also typically observed that CT states exhibit broader emission spectra than locally excited (LE) states (referring to transitions between HOMO and LUMO orbitals with substantial overlap).

In the literature, the prevalent explanation for this spectral broadening is that excitations across an apportioned D-A configuration causes molecular rotation around the D-A connecting bond (Figure 1), resulting in large Stokes shifts and emission spectra with a full-width at half-maximum (FWHM) between 70 and 100 nm.[20] Such broad emission adversely affects color purity, requiring color filters to achieve the color standard requirements when TADF emitters are used in OLED displays.[21,22] The use of color filters is undesirable as it significantly reduces the EQE of OLED displays.

Since the conventional TADF molecular design based on the spatial D-A separation is associated with the combined occurrence of CT and easily excitable torsional rotation between the D-A groups, the question remains whether CT, molecular rotation, or both are responsible for the broad emission characteristics. Rotation between donor and acceptor ensues from the large dihedral angle that is introduced in conventional TADF designs to facilitate twisted intramolecular charge transfer (TICT). To eliminate this potential energy coupling mechanism, molecular interlocking strategies have been developed where added bulky side chains block the rotational motion between donor and acceptor (Table 1).[20,23–28] Assessing whether this approach is indeed effective depends the clarity of the analysis. We first point out that comparing the FWHM in the wavelength vs. energy domains is ambiguous. Because of the scale inversion, the same peak may appear narrower or broader depending on the absolute emission peak position. Presumably, spectral broadening mechanisms involve interactions between electronic and vibronic excitations, and since dispersion of photons and phonons is described by different physical laws, it is preferable to compare conserved quantities such as momenta or energies to better discern the mechanisms that underlie the shaping of the emission line. Interestingly, converting the reported FWHM of some molecules listed in Table 1 into eV dispels the notion that restricting rotation by large steric hindrances would prevent emission broadening. The FWHM on the eV scale does not change, and even a reverse trend is found.

**Table 1. Molecular structures and emission bandwidth of some of the TADF emitters. [a] Measured FWHM in nm, as reported in the references. [b] Converted to eV using the reported emission spectrum in the references**

| | Compounds | Emitter | FWHM (nm)[a] | FWHM (eV)[b] | Ref. |
|---|---|---|---|---|---|
| 1 | 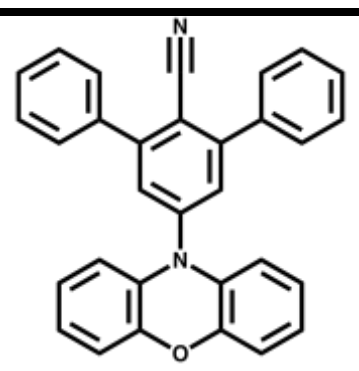 | oPTC | 97 | 0.469 | 20 |
| 2 | 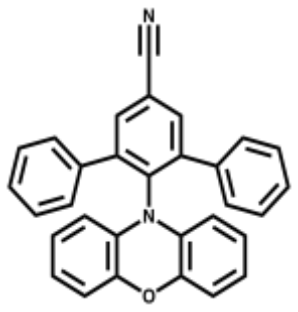 | mPTC | 86 | 0.442 | 20 |
| 3 | 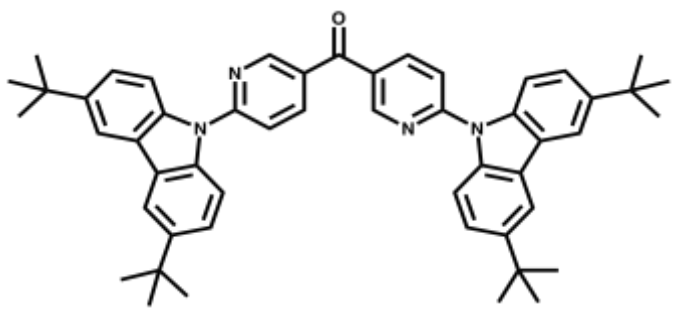 | 3DPyM-pDTC | 62 | 0.347 | 25 |
| **4** | 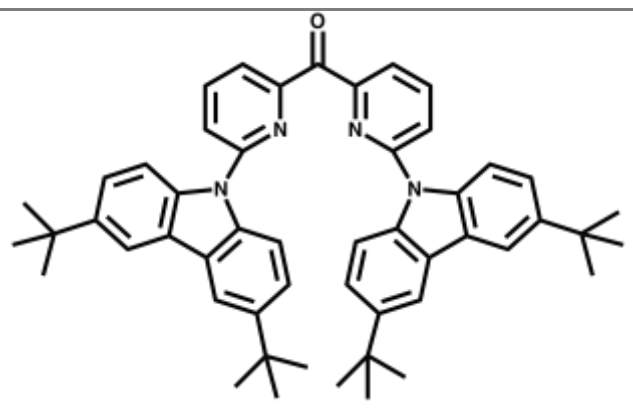 | 2DPyM-mDTC | 89 | 0.420 | [25] |
| **5** | 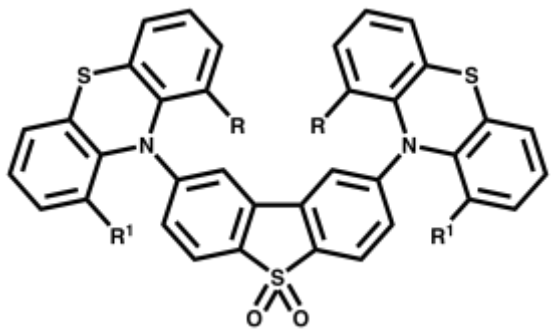 | R=H $R^1$=H Compound1 | 140 | 0.457 | [26] |
| **6** | | R=Me $R^1$=H Compound2 | 130 | 0.477 | [26] |
| **7** | | R=*i*-Pr $R^1$=H Compound3 | 128 | 0.493 | [26] |
| **8** | | R=Me $R^1$=Me Compound5 | 110 | 0.399 | [26] |
| **9** | 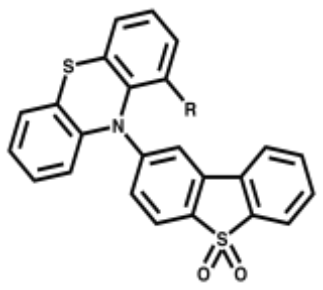 | R=H Compound6 | 100 | 0.391 | [26] |
| **10** | | R=Me Compound7 | 110 | 0.416 | [26] |
| **11** | 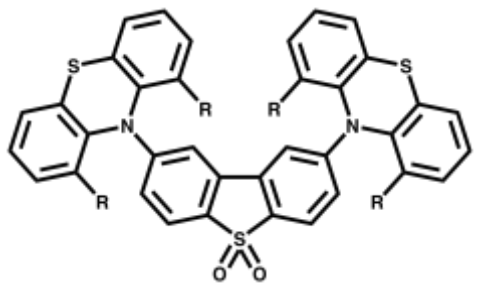 | R=H 1a | 95 | 0.435 | [27] |
| **12** | | R=Me 3a | 100 | 0.481 | [27] |
| **13** | 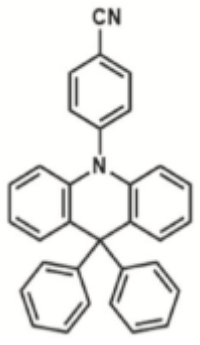 | p-ACN | 65 | 0.433 | [24] |
| **14** | 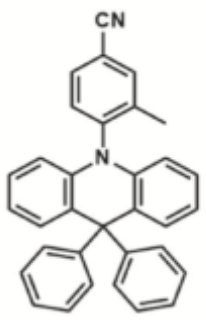 | p-CAN-Me1 | 65 | 0.444 | [24] |
| **15** | 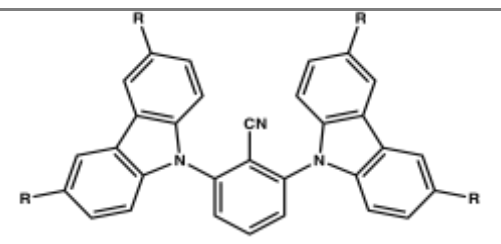 | R=H DCzBN1 | 55 | 0.41 | [23] |
| **16** | | R=Me DCzBN2 | 59 | 0.40 | [23] |
| **17** | | R=t-Bu DCzBN3 | 59 | 0.41 | [23] |
| **18** | | R=MeO | 79 | 0.44 | [23] |
| **19** | 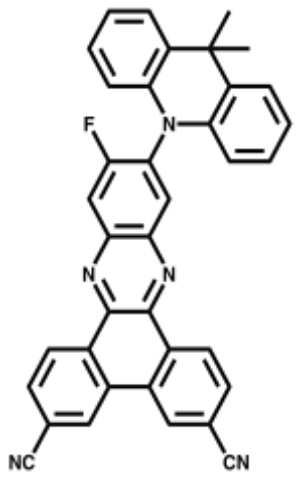 | FBPCNAc | 96 | 0.324 | [28] |
| **20** | 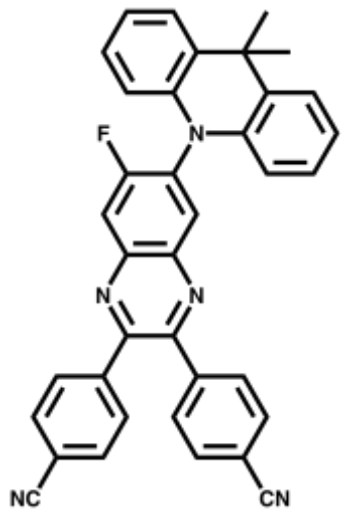 | FDQCNAc | 86 | 0.346 | [28] |

## RESULTS AND DISCUSSION

To identify the key factor that determines the emission bandwidth of TADF emitters, we review the pertinent literature, summarized in Table 1, include several reported compounds in our computational analysis,[20,23] and expand our investigation with a targeted series of new molecules. In 2016, Chen et al. proposed rotational constraint as an effective strategy to improve color purity, based on comparing the emission characteristics of two TADF molecules, mPTC and oPTC (compounds **1** and **2** in Table 1).[20] Of these, mPTC has the larger rotational barrier due to meta-placement of two addition phenyl groups on the benzonitrile acceptor, and it indeed shows a narrower emission band (FWHM = 86 nm) compared to oPTC (FWHM = 97 nm). However, this difference in FWHM is small, especially when expressed in terms of energies, i.e., 0.442 and 0.469 eV for mPTC and oPTC, respectively. Moreover, our calculations show that changing the phenyl substitution position from ortho (oPTC) to meta (mPTC) decreases the charge transfer characteristics of the molecule, since the phenyl groups become less involved in the LUMO orbital, which also results in a narrower emission spectrum. The HOMO-LUMO overlap coefficients for mPTC and oPTC are 0.2531 and 0.1975, respectively. Therefore, mPTC is equally expected to have a narrower emission spectrum based on its weaker charge transfer character. Hence, whether the rotation restriction is really responsible for the narrower emission of mPTC is ambiguous.

The emission band of compound **3** in Table 1 is narrower than that of compound **4**.[25] The authors imply that this is because the rotation between the donor and acceptor group about the C-C bond is prevented for compound **3**. However, they did not provide any computational or experimental proof for this hypothesis. Since CH···N hydrogen bonding is present in both compounds, a comparable restricting effect should ensue. Hence, we expect similar torsional potential energy barriers for the rotation between the donor and acceptor groups in both compounds. In the series of compounds **5-12** in Table 1,[26,27] the emission spectra become narrower as the side chains become bulkier, from hydrogen, methyl, to the isopropyl group. Here as well, the authors attribute the sharper emission spectra to the increased rotational barrier imposed by steric hindrance. However, converting the FWHM from nm to eV reveals that the FWHM are essentially the same, or even the reverse trend is observed. Compounds **13** and **14**[24] have about the same FWHM, even though rotation should be more suppressed in **14,** as an additional methyl group is attached to the acceptor. Compounds **15-18**, in which the R group is H, Me, t-Bu, and MeO, respectively, have the same rotational barrier since the side group does not affect the rotation of the carbazole donor. While the FWHM of these compounds are largely different on the wavelength scale, once convert to energies, the FWHM values are almost the same (0.41, 0.40, 0.41, and 0.44 eV), which provides credence to the choice of energy as the proper spectral scale. Furthermore, our DFT calculations yield similar HOMO/LUMO overlap coefficients for these four molecules (~0.31), essentially unaffected by the extra R(=Me, t-Bu, and MeO) groups. In compound **19**,[28] the two phenyl rings are connected to each other, making it harder for this bulkier group to rotate than in compound **20**, where the phenyl rings can rotate almost freely. Interestingly, the rigidified compound **19** has a broader FWHM than compound **20** (96 nm and 86 nm, respectively). However, converting the FWHM to eV shows that both molecules have about the same emission bandwidth, suggesting that molecular rotation restriction negligibly affects the FWHM. These observations are in line with our hypothesis that charge transfer characteristics dominantly affects the color purity of TADF emitters.

Finally, Hatakeyama et al.[29,30] recently reported organoboron-based planar TADF emitters with small $\Delta E_{ST}$ and excellent color purity. Their design separates HOMO and LUMO based on the multiple resonance effect, without D-A alternation. They exhibit ultrapure TADF color with a FWHM of 28 nm. Following their novel molecular design, Adachi et. al.[31] report another TADF emitter with a FWHM of 34 nm. These sharp TADF emitters all show localized HOMO and LUMO orbitals and their first singlet state ($S_1$) has LE characteristic rather than CT.

To verify our conjecture, we designed and characterized a series of donor-acceptor twisted intramolecular charge transfer (TICT) molecules, in which the charge transfer character and ability to impede intramolecular rotation is systematically varied (Figure 2). Benzonitrile (BN) serves as the acceptor common to all molecules, and the donors include diphenylamine (DPA), carbazole (Cz), phenoxazine (Phe), 9,10-dihydro-9,9-dimethyl-acridine (DMAc), and phenoselenazine (PSeZ).[32,33] oPheBN and mPheBN have previously been investigated in our group,[34–36] and oCzBN and mCzBN have been reported in the literature;[23,37,38] all other molecules are synthesized and characterized for the first time in this work. Using electronic structure calculations, we explore the HOMO and LUMO electron densities and spatial overlap of transition orbitals before and after electron migration. The amount of charge transfer can be assessed based on the degree of spatial overlap between the HOMO and LUMO orbitals. Therefore, to affect the charge transfer we must control the frontier orbitals of a molecule's ground and excited states.

As shown in Figure 3, in general the HOMO and LUMO orbital overlap gradually decreases as the donor strength increases in this series except for oDPABN and oCzBN because diphenyl amine is known to be a stronger donor than carbazole. Even though diphenylamine is a stronger donor, carbazole induces a stronger HOMO and LUMO separation compare to diphenylamine. We believe that disconnected two phenyl rings differently contribute to HOMO electron density than the fused ring donors. Since carbazole is a fused planar ring structure, the dihedral angle between carbazole and the cyanophenyl acceptor is almost 90 degree. This prevents the conjugation of the HOMO orbital from extending to the acceptor, and thereby the HOMO orbital is present mainly on the carbazole group. On the contrary, the two phenyl rings are not connected in diphenylamine, and the HOMO orbital is extended to the acceptor unit, resulting in a larger HOMO and LUMO overlap, despite diphenylamine's stronger donor strength.

Furthermore, we have measured the photoluminescence spectra of oDPABN and oCzBN in various solvents and calculated the degree of red-shifted photoluminescence by solvatochormism. As can be seen in Figures S3 and S4 in supporting information, the emission peak of oDPABN shifts from 420 nm in hexane (less polar solvent) to 458 nm in methanol (more polar solvent), whereas that of oCzBN shifts from 381 nm in hexane to 449 nm in methanol. Accordingly, oCzBN shows a more remarkable red shift with increasing solvent polarity. This is in good agreement with the CT character of the emitting state where a large solvatochromic shift indicates a prominent CT state,[39–42] suggesting a stronger CT character in oCzBN than in oDPABN.

With the least HOMO-LUMO overlap, oPSeZBN has the strongest charge transfer character. The normalized HOMO-LUMO overlap integrals yields a coefficient that quantifies the charge-transfer character of a molecule. These overlap coefficients are listed in Table 2 along with a summary comparison of computational and experimental results. Ultraviolet–visible absorption and photoluminescence spectra of the investigated molecules that were measured in toluene solutions at room temperature are also depicted in the Supporting Information (Figure S1 and S2 Supporting Information).

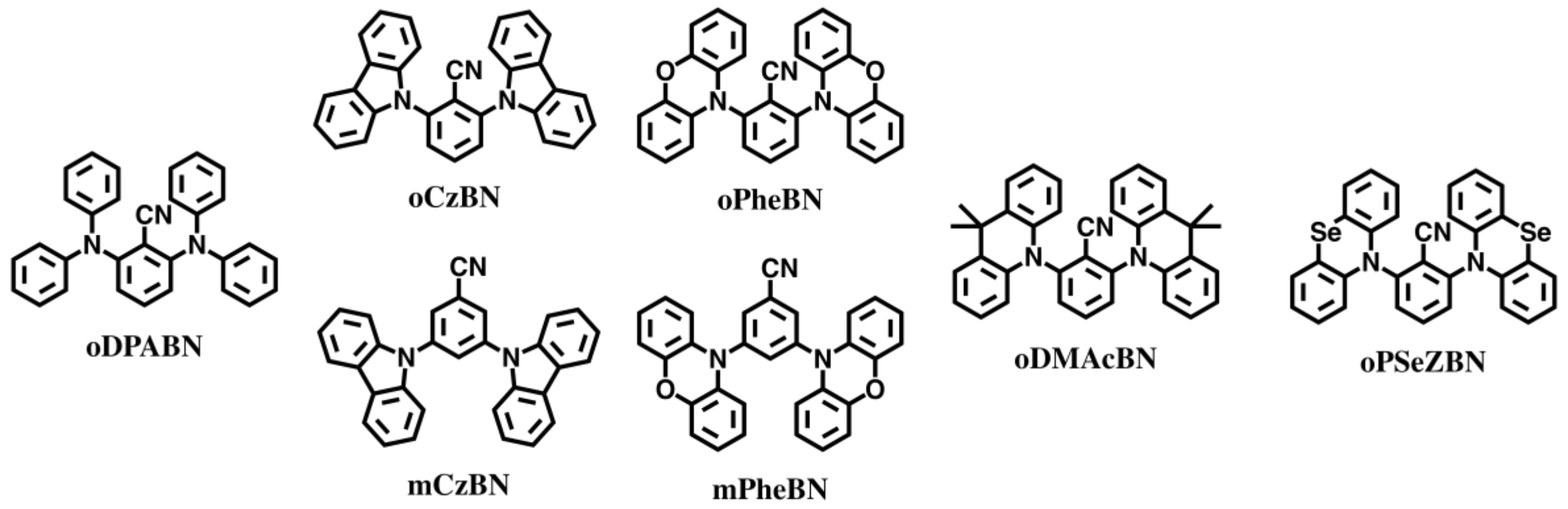

Figure 2. D-A-D TADF molecules investigated in this study.

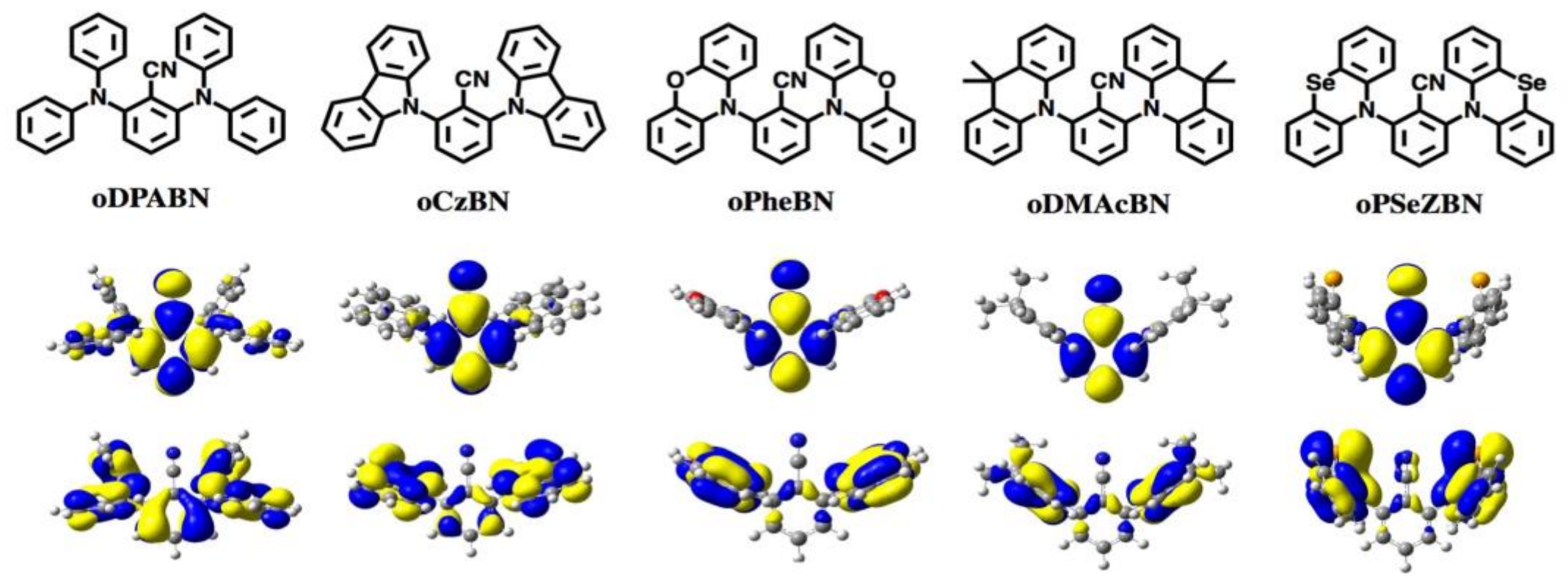

Figure 3. Chemical structures and calculated distributions of HOMOs and LUMOs using CAM-B3LYP/6-31g(d,p) level in vacuum. Stronger donor results in more separated HOMO and LUMO.

**Table 2. Summary of key computational and experimental results. Energy of first singlet ($S_1$), oscillator strength (f) energy of first triplet states ($T_1$), and energy gap between $S_1$ and $T_1$ ($\Delta E_{ST}$) were calculated using DFT at the B3LYP/6-31g(d,p) level in vacuum; overlap integral between HOMO and LUMO were calculated at the CAM-B3LYP/6-31g(d,p) level using multiwfn,[43] the experimental $\Delta E_{ST}$ and emission FWHM were measured in $10^{-5}$ M toluene solution at 25 °C**

| | Compound | Electron Donor | $S_1$[a] (eV) | f[a] | $T_1$[a] (eV) | $\Delta E_{ST}$[a] (eV) | $\Delta E_{ST}$[b, c] (eV) | H/L[a] overlap | FWHM[b] (nm) | FWHM[b] (eV) |
|---|---|---|---|---|---|---|---|---|---|---|
| 1 | oDPABN | Diphenylamine | 3.15 | 0.1863 | 2.81 | 0.34 | 0.313 | 0.4912 | 45.7 | 0.296 |
| 2 | oCzBN | Carbazole | 3.08 | 0.0856 | 2.95 | 0.13 | 0.192 | 0.3097 | 51.8 | 0.381 |
| 3 | oPheBN | Phenoxazine | 2.21 | 0.0000 | 2.19 | 0.02 | 0.00 | 0.2234 | 105.7 | 0.443 |
| 4 | oDMAcBN | Acridine | 2.50 | 0.0000 | 2.48 | 0.02 | 0.00 | 0.2072 | 79.9 | 0.422 |
| 5 | oDPSeZBN | Phenoselenazine | 2.63 | 0.0000 | 2.62 | 0.01 | 0.00 | 0.1795 | 119.1 | 0.473 |

[a] Computational results
[b] Experimental results
[c] $\Delta E_{ST}$ has been measured by comparing the fluorescence and phosphorescence spectra at liquid nitrogen temperature.

As expected, the calculated $\Delta E_{ST}$ decreases as the donor strength increases. Similarly, the lack of orbital overlap, when spatially separating the HOMO and LUMO involved in the excitation reduces mixing of electronic wavefunctions, e.g. $\langle n|\pi^*\rangle$ or CT state, results in a small $\Delta E_{ST}$ but also weak oscillator strength. Therefore, oDPABN, with the largest overlap between the orbitals involved in the $S_1$ transition, has the highest oscillator strength. This value decreases in this series as we increase the donor strength, while the HOMO-LUMO overlap coefficient decreases. The experimental $\Delta E_{ST}$ values are consistent with the calculated $\Delta E_{ST}$, suggesting that our DFT model works well for this system.

Since the $S_1$ transition in these molecules are >97% HOMO->LUMO (Table S2), the HOMO/LUMO overlap coefficient has been used for $S_1$ transition. However, natural transition orbitals (NTOs) are calculated as well by using TD-DFT to analyze the electronic configuration of the $S_1$ state. In Figure S6 and Table S2 one can see HONTO (highest occupied natural transition orbital) and LUNTO (lowest unoccupied natural transition orbital) distributions and overlap extents of the first singlet ($S_1$) state for exciton transformation. The HOMO/LUMO overlap coefficient has also been given for comparison. As evident, the HONTO/LUNTO overlap coefficient is almost the same as HOMO/LUMO overlap. Therefore, the HOMO/LUMO overlap coefficient has been used for the analysis in this work.

Significant emission broadening is observed as the charge-transfer character increases from DPA to PSeZ, thus reducing the overlap between HOMO and LUMO (Figure 3 & Table 2). In this series, the substitution position remains unchanged; only the donor strength is varied, which is accompanied by a change in dihedral angle. As evident in Figure 4, we observe a linear correlation between FWHM and the HOMO-LUMO overlap coefficient.

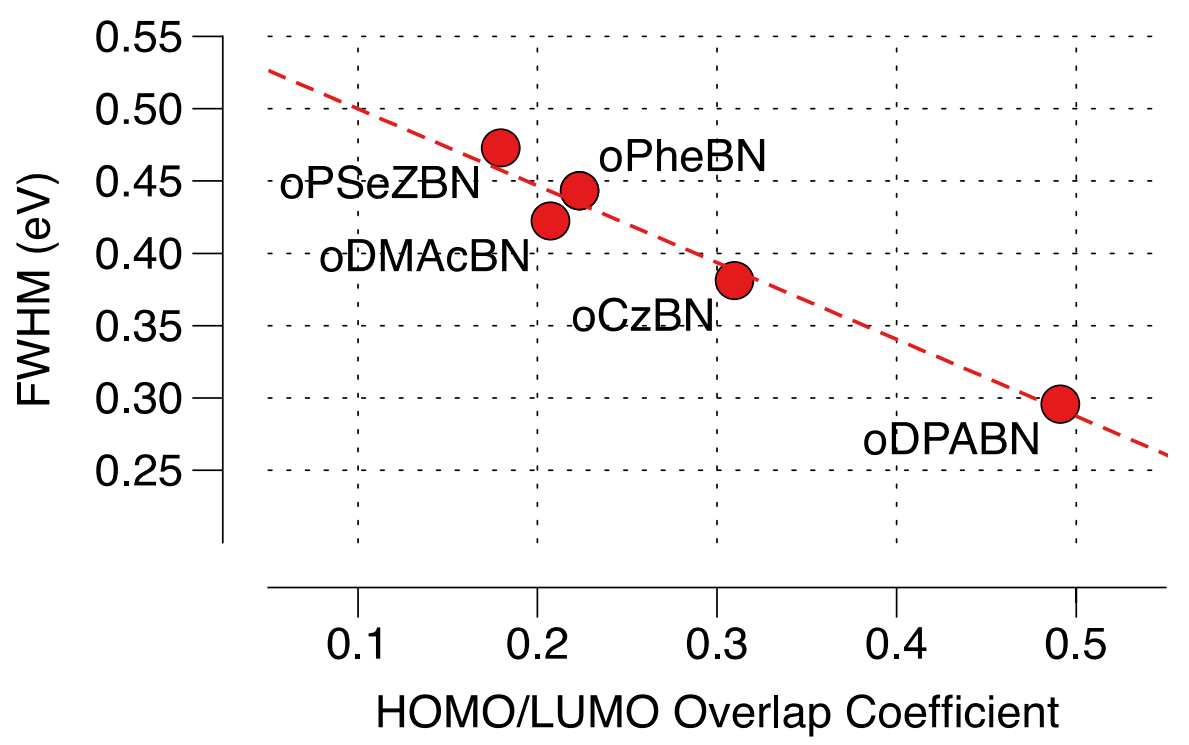

Figure 4. Correlation between measured FHWM and calculated HOMO/LUMO overlap.

If rotation of the donor group made the emission broader, one would expect that oDPABN has a broader bandwidth than oCzBN. In the latter, the two phenyl rings in the carbazole unit are connected to each other, making it harder for this bulkier carbazole to rotate. By contrast, the phenyl rings on the diphenylamine group can rotate almost freely, as substantiated by the low torsional potential energy barrier shown in Figure 5. However, the TADF emission spectrum of oDPABN is much sharper than that of oCzBN; 0.296 eV compared to 0.381 eV. oCzBN has stronger charge transfer character (lower HOMO-LUMO overlap, see Table 2), which results in a broader emission bandwidth compared to oDPABN. This is strong evidence that charge transfer character controls the emission bandwidth of TADF compounds, while rotation has a negligible effect.

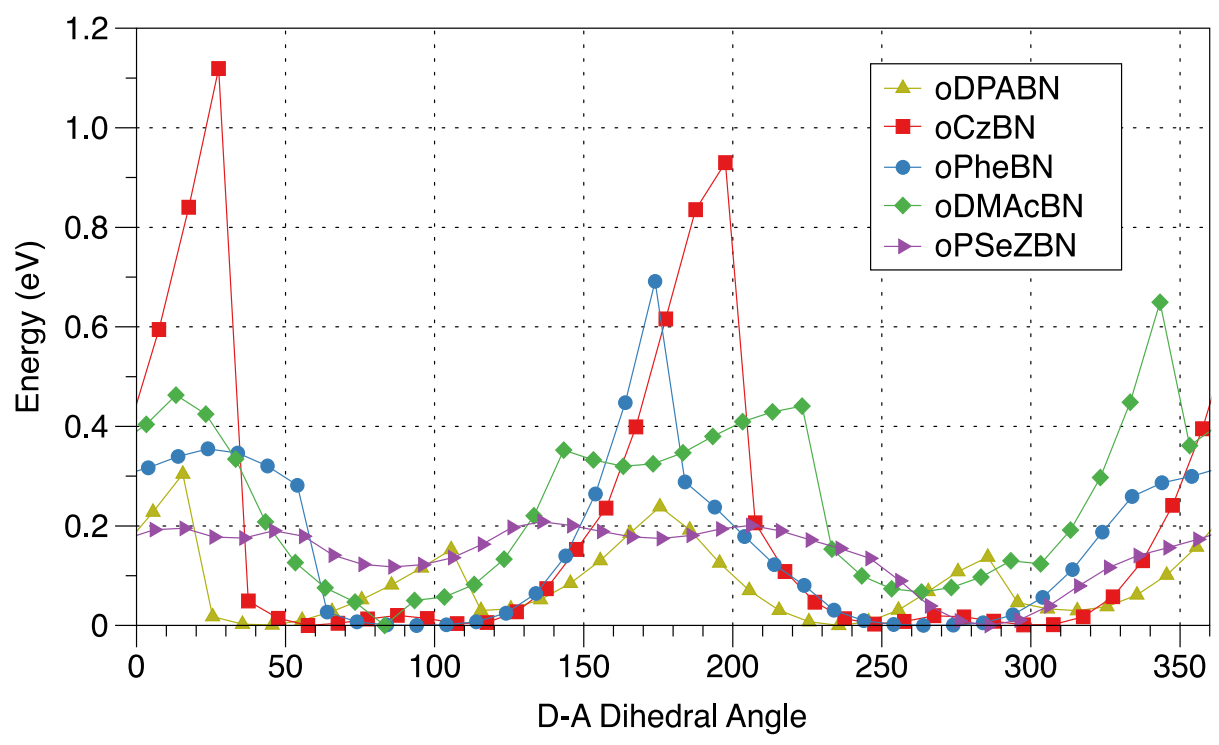

Figure 5. Torsional potential energy of investigated molecules (ortho position) as a function of donor-acceptor dihedral angle from DFT calculations at the CAM-B3LYP/6-31g(d,p) level.

We also synthesized the meta version of the carbazole (mCzBN) and phenoxazine (mPheBN) compounds. The meta position is farther away from the cyano group than the ortho position. Ortho and meta substituted molecules have about the same charge transfer characteristic, since their donor strength is the same; only the position of the donor is different. We analyzed the optimized molecular structures in both $S_0$ and $S_1$ states, using quantum mechanical calculations (Figure 6(a)). As designed, oPheBN and mPheBN adopt nearly orthogonal D-A-D conformations between the Phe and the BN units, with dihedral angles of 93.9° and 99.6°, respectively, in the $S_0$ state, while these dihedral angles are 87.3° and 84.1° in the $S_1$ state. With 6.6° compared to 15.5°, geometric differences between the ground and excited states are smaller for oPheBN than for mPheBN. Furthermore, quantum mechanical calculations reveal that mPheBN has a much lower energy barrier for torsional motion between donor (Phe) and acceptor (BN) than oPheBN (Figure 6(b)). Accordingly, oPheBN is rigid upon photoexcition due to the steric hindrance imposed by the cyano group. Conversely, mPheBN is more flexible with the Phe groups at meta positions, where their rotation is less encumbered by the cyano group. The same behavior is observed for oCzBN and mCzBN (Figure 6(b)). The torsional potential barrier for carbazole compounds are higher than phenoxazine because the carbazole group has a five-membered ring that cannot bend out of plane, while the central ring in phenoxazine bends to lower the molecule's energy when the dihedral angle is raised.

The HOMO-LUMO overlap coefficents for oPheBn and mPheBN are 0.2234 and 0.2304, which is why they have similar charge transfer character. Despite the different rotational restriction caused by the cyano goup in oPheBN, both exhibit about the same FWHM: 0.443 eV for oPheBN and 0.459 eV for mPheBN. Also, they both exhibit minimal emission broadening (<0.005 eV) upon thermally activating rotation the temperature range between 5°C and 80°C (Figure 6(c)). Since the rotational barrier is small for mPheBN, increasing the temperature should provide the Phe group with enough energy to overcome the rotational barrier and broaden the emission. However, this is not the case. oCzBN and mCzBN exhibit the same trend (Figure 6(b,c)). The torsional barrier for oCzBN is higher than for mCzBN, because of the interference with the cyano group. Similar to the phenoxazines, both show minimal emission broadening upon raising the temperature. This strongly supports that D-A rotation negligibly affects the emission bandwidth of TADF compounds, while charge transfer plays the major role.

We also measured the steady state photoluminescence spectra of oCzBN in various solvents ranging from hexane (lowest polarity) to methanol (highest polarity). Positive solvatochromism has been observed as expected. The emission peak red shifts from 381 nm in hexane to 449 in methanol (Figure S4(a)). Significant emission broadenning is observed as the polarity of the solvent increases. We have also converted the abscissa unit from nm to eV (Figure S4(b)). Converting the FWHM to eV reveals that the FWHM still gets broader as the solvents polarity increases (Figure S5). In other words, the same compound, oCzBN, shows different FWHM depending on the solvent choice. We used the polarizible continuum model (PCM) at CAM-B3LYP/6-31G(d,p) to incorporate the solvent effect. As it is evident in Table S1, our DFT results show that the HOMO-LUMO overlap coefficient of oCzBN decreases as the polartity of the solvent increases. This is exactly the same trend that we observed from ortho substituted molecules in Figure 4. However, we believe that the PCM model underestimates the solvent effect and changing the polarity of solvent would have resulted in a noteworthy change in the HOMO-LUMO overlap coefficients, making the overlap coefficients of oCzBN in various solvents are close to each other.

Having observed the drastic emission broadening by increasing charge transfer character, and that restricting molecular rotational has a minimal effect on the color purity of TADF emitters, LE transitions seem to be the only effective way to improve the color purity. However, molecules with LE emitter states tend to have rather large $\Delta E_{ST}$, making up-conversion challenging. As a result, TADF becomes a matter of optimizing molecular designs so that $\Delta E_{ST}$ is small enough for efficient TADF and the emission is still sharp.

Separating HOMO and LUMO via multiple resonance effect, suggested by Hatakeyama et al.,[29,30] which results in an LE state, is so far the most promising route for high color purity TADF emission. The molecular resonance TADF molecules presented by Hatakeyama et al, DABNA1 and DABNA2, have the HOMO/LUMO overlap coefficient of 0.6038 and 0.5869, respectively (calculation done in this work; using B3LYP/6-31G(d,p)), implying their LE characteristics. As a matter of fact, they have a higher LE characteristic than TADF emitters with weak donors (like oDPABN), narrowing FWHM as small as 14 nm. However, the carbon-boron bond in these compounds causes poor stability, resulting in significant roll-off in OLED applications. Moreover, the work by Dias et. al.,[44] has shown that even with $\Delta E_{ST}$ as large as 0.3 eV, 100% TADF efficiency is achievable by introducing a $^3n\pi^*$ triplet state. Also, Higginbotham et. al.[45] were able to increase the rISC rate to a near 2 orders of magnitude by mixing the $n\pi^*$ with the donor $\pi$ orbitals, whereas the singlet-triplet splitting, $\Delta E_{ST}$, remains unaffected. Therefore, future efforts are directed towards designing TADF molecules with weak donors, resulting in higher HOMO-LUMO overlap and a narrower emission bandwidth. A presence of the $^3n\pi^*$ triplet state is necessary to enhance rISC and compensate for the relatively large singlet-triplet energy gap.

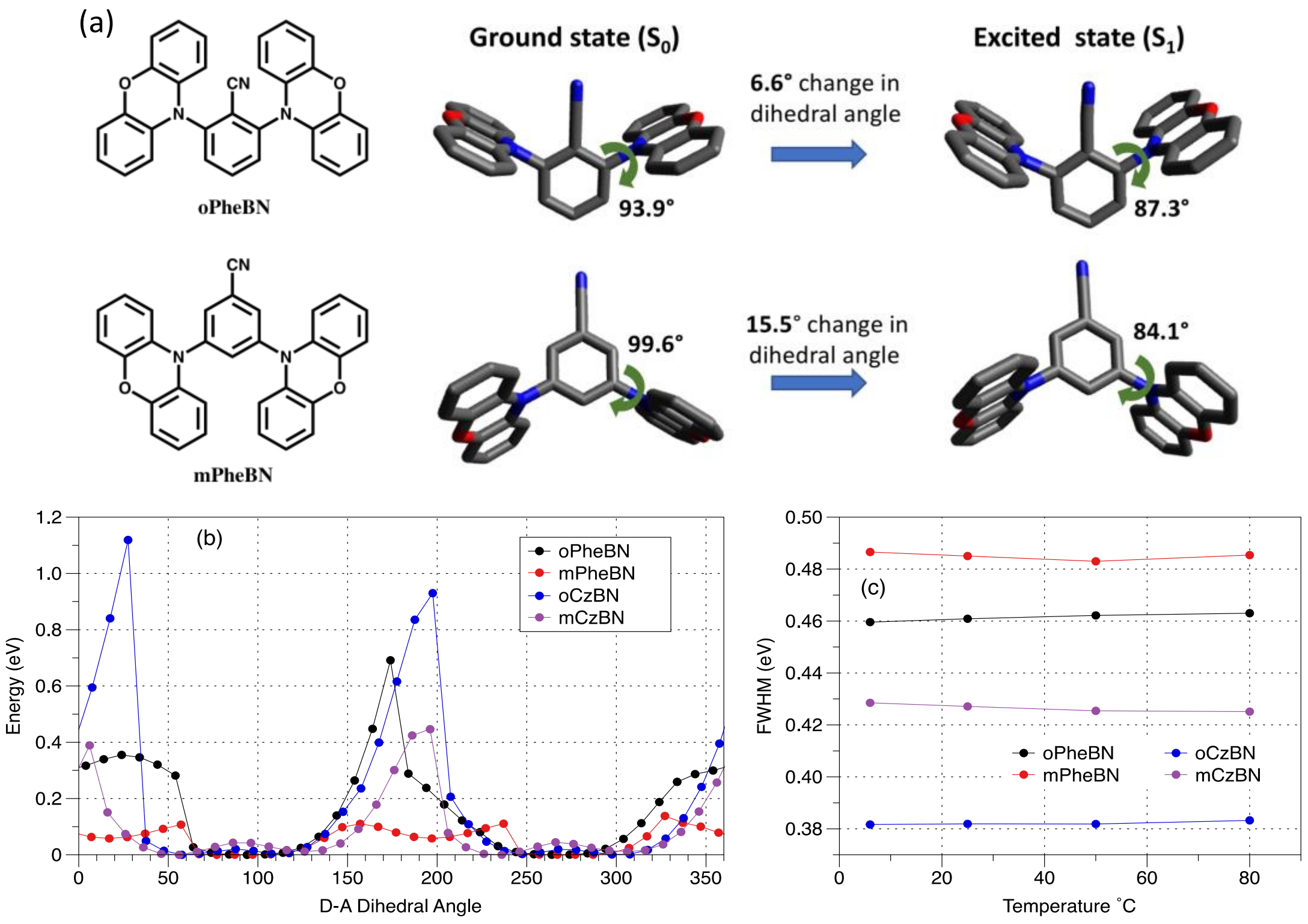

Figure 6. (a) Chemical structures of oPheBN and mPheBN and the differences in the differences in the optimized geometries between the ground and first singlet excited states, hydrogens are not shown for clarity. oPheBN shows 6.6° change in the dihedral angles between the donor and acceptor, however, mPheBN shows a higher dihedral angle change of 15.5°. (b) Torsional potential energy of of oPheBN (black), mPheBN (red), oCzBN (blue), and mCzBN (purple) as a function of donor-acceptor dihedral angle from DFT calculations at the CAM-B3LYP/6-31G(d,p) level. (c) Temperature dependence FWHM of oPheBN (black), mPheBN (red), oCzBN (blue), and mCzBN (purple) in $10^{-5}$ M toluene solution (The FWHM discrepancy between (c) and table 1 is because different fluorimeters were used, but measurements are consistent within each data set).

## CONCLUSIONS

In summary, we designed and synthesized a series of D-A-D type TADF emitters and altered the charge transfer characteristics and rotational restrictions in these molecules. Compounds with the same rotational barrier show a large increase in FWHM when the charge transfer character gets stronger. However, even with increased rotational restrictions caused by the cyano group, emitters showed minimal change in their FWHM. If increased restriction of the molecular rotation were effective in improving color purity, one would expect broader emission spectra upon increasing the temperature, since the molecules gain energy to overcome the rotation barrier. However, temperature has minimal effect on color purity. These results, along with examining data for various compounds described in the literature, strongly suggest that molecular rotation restriction negligibly affects the FWHM and that the intrinsic TADF emission bandwidth is mainly controlled by the charge transfer character. Our combined experimental and computational results provide insightful understanding about rational molecular design of TADF molecules with a narrow emission spectrum.

## ASSOCIATED CONTENT

### Supporting Information

The Supporting Information is available free of charge on the ACS Publications website.

> Synthetic routes of the molecules investigated in this work; UV–vis absorption and Photoluminescence spectra; Synthesis; Computational method; and Cartesian Coordinates

## AUTHOR INFORMATION

### Corresponding Author

*E-mail: John Kieffer kieffer@umich.edu

*E-mail: Jinsang Kim jinsang@umich.edu

### Notes

The authors declare no competing financial interest.

## ACKNOWLEDGMENT

This research was supported by funding from the US National Science Foundation, grant number **NSF DMR-1435965**. The compounds oPheBN and mPheBN were previously synthesized in our group by Dr. Mounggon Kim, and used for the measurements in this work.